# The universe would not be perfect without randomness: a quantum physicist's reading of Aquinas.

**Valerio Scarani**

Centre for Quantum Technologies and Department of Physics, National University of Singapore, Singapore 117543

**Abstract**

Randomness is an unavoidable notion in discussing quantum physics, and this may trigger the curiosity to know more of its cultural history. This text is an invitation to explore the position on the matter of Thomas Aquinas, one of the most prominent philosophers and theologians of the European Middle Ages.

**Introduction**

Physical determinism has been a powerful methodological assumption in science since the dawn of the modern era. It is so engrained in our culture, that the sentence "to find a scientific explanation" usually means "to point to an antecedent situation, which existed (or can reasonably be assumed to have existed) in the past, from which the present situation follows by law of necessity". Since the first half of the 20th century, physical determinism has been famously challenged by quantum physics, with the theorem of Bell that we are celebrating here as one of the milestones. But neither fifty years of a theorem, nor a century of quantum physics, can easily dispose of an intellectual option that dates back at least to the time of Democritus and Lucretius, and has deeply shaped the last few centuries of human thought. The urge for physical determinism explains why many laymen are still associating quantum physics with some esoteric quest rather than with the most successful scientific endeavor of all times, popular journals thriving on the misconception by suggesting that the latest research paper may be the one lifting the veil. Determinism appeals to the specialists too[1]: for those who support Bohmian mechanics or the many-worlds interpretation, determinism is a cornerstone that should not be removed; one should rather abandon other supposed features of the physical world and of our assessment of it.

Surely, chance and randomness in their various meanings have also had their supporters in the recent scientific and philosophical debates. Nevertheless, I was intrigued when I chanced over the following statement in Thomas Aquinas' *Summa contra gentiles* (CG), most probably written around the years 1260-64:

---

[1] For a recent perspective on determinism by a quantum colleague: L. Vaidman, Quantum theory and determinism, http://arxiv.org/abs/1405.4222.

> **"it would be contrary to [...] the perfection of things, if there would be no chance events".**
> **(CG, Book 3, Chapter 74)**

Which arguments in favor of chance as a feature of nature could Aquinas bring up, several centuries prior to Darwinian evolution, deterministic chaos and quantum physics – and obviously without any hint thereof?

I decided to write about this text and a few related passages by the same author[2] for these proceedings. I knew it would be a challenge: in the words of a philosopher friend of mine, I opened a window and discovered an ocean, for whose exploration I am not well equipped. The purpose of this text is not to claim a place among the explorers, but to invite others to come at the window[3].

**Aquinas on God's Providence**

The third book of the CG is devoted to "Providence", which is *God's governance of creation*. As an orthodox Christian philosopher and theologian, Aquinas had to juggle to accommodate both an all-powerful God and really free human beings. Alongside this extremely important anthropological question, certainly still debated today[4], comes a more neutral but general observation of nature: all the beings that we perceive are limited in their being. Aquinas integrates both considerations in a response that rings surprisingly modern: not only "human free will", but *the whole creation, including its material aspect, possesses a relative autonomy from God*. This autonomy is going to be the foundation for the discussion of "fortune and chance".

If the reader is yawning at the apparent banality of this observation, they have better wake up quickly because the statement is not trivial at all. In fact, if prompted to define "autonomy from God", we tend to refer to the *cosmology*[5] of the time of Leibniz and Newton. God's role is that of a watchmaker who builds a perfect mechanism, which later evolves "autonomously" – meaning here, according to the deterministic laws of classical physics. The autonomy of the watch from the watchmaker is indeed the fitting setting for physical determinism.

Aquinas lived at a time where the watch had not been invented yet. The cosmology of his time is twice removed from us, transiting as we are to a new

---

[2] Interestingly, a quantum colleague, who contrary to me cannot be suspected of Catholic leaning, has also advocated recently a re-discovery of Aquinas' thought: D.M. Appleby, Mind and Matter, http://arxiv.org/abs/1305.7381. This text sketches a much more ambitious program than mine here; Aquinas' philosophy is proposed as a possible way of avoiding Cartesian dualism.

[3] A very convenient summary of Aquinas' philosophical work is available as: Aquinas, *Selected philosophical writings* (Oxford World's Classics, Oxford University Press, 2008). The Latin text and English translation of most of Aquinas works can be found online in http://www.dhspriory.org/thomas/. When available, I'll give the link to this website.

[4] See e.g. http://plato.stanford.edu/entries/providence-divine/; Aquinas' position is discussed in paragraph 6.

[5] In this text, "cosmology" will be used in the sense of *Weltanschauung*, not in the sense of the discipline of physics that studies the universe at large.

cosmology of evolutionary flavor. Aquinas' metaphor for the universe is that of an orderly kingdom[6]:

> "Political life offers a parallel: for all the members of the household are ordered to one another by subordination to the master of the house, and then that master and all other masters of households in a city ordered to one another and to the ruler of the city; and he with all his fellows in a kingdom, ordered to the king".
>
> (CG 3, 98)

This cosmology contains two elements that are currently banned from much of the intellectual discourse on nature. These are *finality* and the existence of a *hierarchy of beings*.

Concerning *finality*, or finalism, I feel incapable of providing an even moderately competent discussion[7]; however, there is a point that I want to bring to the attention of the reader. The absence of finalism was a defining feature of the atomism of Democritus and Leucippus, as noticed explicitly by Aristotle. Early in his career, Aquinas writes about this[8]:

> "First, we have to know that some stated that there is no providence for anything, that everything happens by chance: this was the position of Democritus and the other ancient authorities who denied agent causes and affirmed only material causes. But this position has been sufficiently refuted in philosophy".
>
> (*Scriptum super Liber Sententiarum* I, d.39)

A doctrine that "affirms only material causes": to our ears, this would make of Democritus a precursor of determinism, and as such he is indeed presented in philosophy textbooks. Aquinas' highlights rather the fact that, in such a doctrine, "everything happens by chance", that is "without any providence" – with a more modern twist, we may say "without meaning". Atomism and determinism have gone hand-in-hand for centuries – till quantum mechanics cast its shadow on the idyll. Is it a full-scale betrayal or a test that will further consolidate the relationship? This is what we are still debating.

---

[6] http://www.dhspriory.org/thomas/ContraGentiles3b.htm#98. I am citing from the more readable translation of Timothy McDermott in the Oxford book cited above.

[7] For a text that mentions Aquinas extensively and provides a glimpse of the complexity of the issue, see http://inters.org/finalism, paragraphs I and II.

[8] I am grateful to V. Cordonier for sharing with me her text "La doctrine aristotélicienne de la Providence divine selon Thomas d'Aquin" [in: P. D'Hoine, G. van Riel (ed.), Fate, Providence and Moral Responsibility in Ancient, Medieval and Early Modern Thought. Studies in Honor of Carlos Steel (Peters, Leuven, 2014) pp. 495-515], in which I found this very relevant citation (footnote 14). My translation.

Let us now discuss the *hierarchy of beings*. In his theory of knowledge, Aquinas considers "being" as the first we grasp[9]: that is, before knowing that it is (say) a tree, or that it is in that place, or that it is good for us, we know that it "is". From the perspective of our knowledge, thus, the existence of various beings is not a truth to be derived, but the starting point. In CG, Aquinas rather takes God's creation as starting point, and argues why such a creation must consist of multiple finite beings[10]:

> "Now, created things must fall short of the full goodness of God, so, in order that things may reflect that goodness more perfectly, there had to be variety in things, so that what one thing could not express perfectly could be more perfectly expressed in various ways by a variety of things. [...] And this also draws attention to how great God's perfection is: for the perfect goodness that exists one and unbroken in God can exist in creatures only in a multitude of fragmented ways".
>
> (CG 3, 97)

My last introductory comments will be on the following citation from the same chapter:

> "Clearly then, the dispositions of providence have their reasons, but reasons that presuppose God's will. All this allows us to avoid two kinds of mistakes. First, the mistake of those who believe everything comes from the simple will, devoid of reason [...]; and secondly, the mistake of those who say the causal order is a necessary consequence of God's providence".
>
> (CG 3, 97)

The first mistake, that Aquinas is convinced of having disposed of, is the belief that there is no rationality in the world, that everything is pure arbitrariness from God's decisions[11]. The second mistake addresses our concern. It is beyond human capacity to assess the purpose of God[12], which is the ultimate final cause;

---

[9] It is acknowledged that this is a foundational element of Aquinas' theory of knowledge. If prompted to find a citation, probably the most famous is *primum enim quod in intellectum cadit, est ens* (*De Pot.* 9, 7, ad 15 http://www.dhspriory.org/thomas/QDdePotentia9.htm#9:7).

[10] http://www.dhspriory.org/thomas/ContraGentiles3b.htm#97; I am citing again the translation from McDermott.

[11] Aquinas was aware that some Arabic philosophical schools of Andalusia promoted this doctrine. One century after Aquinas' death, it was going to be championed again, this time in the Christian world by William of Ockham. It still lurks behind many anti-scientific attitudes of our times.

[12] Aquinas inherited the "negative theology" of Pseudo-Dionysius: we can't know anything of God's plan, besides what He chooses to reveal to us. The Revelation accepted by both deals with the finality set by God for human beings, but says close to nothing about that for the material world (indeed, basically all that Christian belief has to say on the final destiny of matter is that the final destiny of humans does involve a material element, in the following of Jesus, who resurrected with what we could call an "upgraded version" of his own body).

but whatever it is, it does not determine a unique unfolding[13] because created beings can act as causes, including final (i.e. they can act for some purpose that is *their* purpose, not God's). This is the "relative autonomy" that was mentioned. Some beings being more perfect than others, their causality is also of different power: in particular, it may fail to produce the desired effect. This brings us to the main objective: a discussion of the case for "fortune and chance".

**The case for fortune and chance**

In our main reference, CG 3, 74, Aquinas lists five arguments to defend the thesis that fortune and chance are compatible with God's providence[14]. I rephrase them in my words:

1. If nothing rare would happen, we would conclude to necessity. Thus "fortune and chance" are the manifestation of contingency, which is God's respect of the autonomy of created beings.
2. The second argument combines finality and finiteness: all beings act for an end, but finite beings may fail with regards to the intended end, thus bringing about unintended effects.
3. The third argument is different: it is the classic *concursus causarum*. Since God does not determine everything and each being has its own autonomy, it is possible that initially independent causal chains collide to produce an unexpected effect. The example of Aquinas is more than clear: "For example, the discovery of a debtor, by a man who has gone to market to sell something, happens because the debtor also went to market".
4. In yet another chance of perspective, the finite beings are no longer considered as agents, but as beings, whose properties are not all necessary. The actual text, a scholastic demonstration, sounds very convoluted to us; so let me try my own example. A given woman is a human being, is tall, is dressed in blue, and is a physicist. While being human" is obviously essential, the other features look accidental and uncorrelated among them – but who knows, maybe there is a deep common cause for all the features of this woman? Aquinas argues that against such a higher causality: it is proper of finite beings to have indeed many accidental features.
5. The fifth argument has some flavor of the second and the third: the power of a finite cause is necessarily finite and therefore cannot extend to all things that can happen.

A comprehensive commentary is beyond my capacity, since I have not dived into the ocean. But there is another island visible from the window, which is impossible to ignore. Indeed, fortune and chance had already been discussed by Aristotle in his *Physics*, Book II, Chapters 4-6. After having introduced his classification of four causes, Aristotle discusses the opinion that fortune, chance or "spontaneity" are also causes; he reasons that all these are indeed real, but are not proper causes, thus justifying his previous classification. More or less in the same years[15] as CG, Aquinas wrote a commentary to this work, the *Commentaria* [or *Expositio*] *in Octo Libris Physicorum*. A scholastic commentary was in fact a series of lectures reviewing the text of an authority point by point. It is not easy to follow, insofar as the commentator can be critical of one statement and will nevertheless go ahead with the commentary of the next. I just want to point to a few points that shed some light on the text from CG.

---

[13] Notice again the discrepancy with the later cosmology: Liebniz argued that the God-watchmaker must have created the best possible world in all the details of its gears.
[14] http://www.dhspriory.org/thomas/ContraGentiles3a.htm#74

- Even if it comes unexpectedly last, I want to mention first that "chance" and "fortune" are defined in Aristotle's chapter 6, commented by Aquinas in his 10th lecture[16]. Fortune (misfortune) implies happiness (sadness) and thus is proper of beings that can experience happiness; whereas chance is a neutral word that applies to all beings. The fact that these definitions are elaborations on the common meaning of those words may explain why Aquinas did not define those terms in CG.
- The example of the two men meeting by chance in the market, which we saw in argument #3 above, comes directly from Chapter 4 of Physics. As stressed by Aquinas[17], the example here is meant to show that "fortune" is certainly not always a cause: indeed, here one would speak of fortune (especially for the creditor), but the cause of each person going to the market was not "fortune", it was "to buy something".
- At the beginning of chapter 5 of *Physics*, Aristotle makes what we may call nowadays call a "phenomenological study" of the cases in which chance or fortune are invoked as causes. It opens with an observation similar to the one in argument #1 above: one speaks of chance when things happen rarely. In his Lecture 8[18], Aquinas writes: "it seems that this division [in things that happen always, frequently or rarely] of the Philosopher is insufficient, for there are some happenings which are indeterminate". From what I understand, "indeterminate happenings" (*contingentia ad utrumlibet*) refers to events whose frequency cannot even be defined.

**Message in an old bottle**

In these notes from the window, I tried to grasp Aquinas' effort of rationalization. It's a message in a bottle from a cultural world that is no longer ours: in particular it would be grossly anachronistic to read Aquinas as a precursor of quantum physics[19]. But we are allowed to read the message and derive some inspirations for our times.

---

[15] The most probable date is around 1268, which would put it some five years after the most probable date for CG – anyway, all that matters here is that the two texts belong to the same period, so that in first approximation we can assume them to be consistent with each other.
[16] *In Physic.* II, l.10 (http://www.dhspriory.org/thomas/Physics2.htm#10).
[17] *In Physic.* II, l.7 (http://www.dhspriory.org/thomas/Physics2.htm#7).
[18] *In Physic.* II, l.8 (http://www.dhspriory.org/thomas/Physics2.htm#8).
[19] Authors like Heisenberg, Jauch and Piron, have used the wording of "potency and act" in their attempts to appraise quantum physics. Inspired by this, some years ago I browsed extensively Aquinas' works to see if a hint of the quantum could be found there: I can say with high confidence that such is not the case. Let me give an example. For Aquinas (and Aristotle), the statement "I am in potency of being at B" means that I actually am at A, and by motion I could exchange my "being at A" with "being at B". In no way they had thought of "not being actually

Aquinas' study is very far from a naive god-of-the-gaps argument, which would run: "there are things I cannot predict, the only possible explanation is to invoke the intervention of God or some other spirit". Chance and fortune are neither God's doing nor the devil's: they are the manifestation of the finiteness of created beings, and of the autonomy that God's providence gave them. Since this autonomy is a sign of God's respect for his creation, chance and fortune are to be considered positive realities.

I want to stress that the existence of God is not an assumption for the argument. As we said, Aquinas does believe in an all-powerful God, and his challenge was to present a doctrine of providence that does not end up in determinism. One could make a God-free case for randomness along similar lines, as long as one accepts the existence of finite autonomous beings. This is far from universally accepted: many philosophies and mysticisms around the world and across the centuries have upheld the doctrine that behind the appearance of a multitude of beings there is only one Being[20]. Such "holistic" or "pantheistic" doctrines may have their own way to deal with freedom and chance, of which I know little, but certainly they won't follow Aquinas' path.

I want to conclude on a recollection from the conference, whose proceedings you are reading. I happened to give two talks there: the first one was my own, the second one was the one of Nicolas Gisin, who had been retained in Geneva by urgent family matters. This second talk lead to a broad discussion, during which Anton Zeilinger dropped the suggestion that a better appreciation of quantum physics may pass through a rediscovery of *finality*. There and then, I thought of Aquinas. Now, after reading CG 3, 74, it's easy to recall that Zeilinger and Brukner have also promoted the explanation of quantum randomness as a consequence of the *finiteness* of the information that can be stored in a quantum system. Surely, if all nature cares is that some statistics are respected, the concrete way to get there may be left to "chance". Happy as I am with this argument, I feel it does not explain why nature wanted the statistics to violate Bell inequalities. Maybe God wanted us humans to be able to *certify* intrinsic randomness?

**Acknowledgements**
This text has greatly benefitted from discussions with and feedback or encouragement from: Antonio Acín, Michael Brooks, Jeff Bub, Jonathan Chua Yi, Valérie Cordonier, Artur Ekert, Berge Englert, Nicolas Gisin, Matthew Leifer,

---

localized anywhere", which is what Heisenberg, Jauch and Piron were aiming at – as to whether this extension is legitimate and successful, I am skeptical but with no strong feelings.

[20] When speaking of "holism" or "pantheism", Indian-born religions come immediately to mind, but similar hints can be found even in Plato. Very relevant for our story is the fact that Averroes explicitly commented Aristotle in a holistic sense, making him very suspicious in the Christian world: the "redemption" of Aristotle from that interpretation was arguably Aquinas' greatest challenge (see e.g. G.K. Chesterton, *Saint Thomas Aquinas, the "Dumb Ox"*, several editions). In later times, some humanists will promote again the doctrine of an *anima mundi*, and Spinoza will champion a renewed form of pantheism. Presumably some members of the Church of the Larger Hilbert Spaces have a similar doctrine in mind.